
\input epsf.tex

\def\pmb#1{\setbox0=\hbox{#1}%
  \hbox{\kern-.025em\copy0\kern-\wd0
  \kern.05em\copy0\kern-\wd0
  \kern-0.025em\raise.0433em\box0} }

\catcode`@=11
\def\leftrightarrowfill{$\m@th\mathord\leftarrow \mkern-6mu
  \cleaders\hbox{$\mkern-2mu \mathord- \mkern-2mu$}\hfill
  \mkern-6mu \mathord\rightarrow$}
\def\overleftrightarrow#1{\vbox{\ialign{##\crcr
     \leftrightarrowfill\crcr\noalign{\kern-1pt\nointerlineskip}
     $\hfil\displaystyle{#1}\hfil$\crcr}}}
\catcode`@=12

\def\approxge{\hbox {\hfil\raise .4ex\hbox{$>$}\kern-.75 em
\lower .7ex\hbox{$\sim$}\hfil}}
\def\approxle{\hbox {\hfil\raise .4ex\hbox{$<$}\kern-.75 em
\lower .7ex\hbox{$\sim$}\hfil}}

\def \abstract#1 {\vskip 0.5truecm\sepline\vskip 0.5truecm
$$\vbox{\hsize=15truecm\noindent #1}$$}
\def \SISSA#1#2 {\vfil\vfil\centerline{Ref. S.I.S.S.A. #1 CM (#2)}}
\def \PACS#1 {\vfil\line{\hfil\hbox to 15truecm{PACS numbers: #1 \hfil}\hfil}}

\def \hfigure
     #1#2#3       {\midinsert \vskip #2 truecm $$\vbox{\hsize=14.5truecm
             \seven\baselineskip=10pt\noindent {\bcp \noindent Figure  #1}.
                   #3 } $$ \vskip -20pt \endinsert }

\def \hfiglin
     #1#2#3       {\midinsert \vskip #2 truecm $$\vbox{\hsize=14.5truecm
              \seven\baselineskip=10pt\noindent {\bcp \hfil\noindent
                   Figure  #1}. #3 \hfil} $$ \vskip -20pt \endinsert }

\def \vfigure
     #1#2#3#4     {\dimen0=\hsize \advance\dimen0 by -#3truecm
                   \midinsert \vbox to #2truecm{ \seven
                   \parshape=1 #3truecm \dimen0 \baselineskip=10pt \vfill
                   \noindent{\bcp Figure #1} \pretolerance=6500#4 \vfill }
                   \endinsert }

%
\def \ref
     #1#2         {\smallskip \item{[#1]}#2}
\def \sepline     {\medskip\centerline{\vbox{\hrule width5truecm}} \medskip}

\def \tabrul2     {\noalign{\vskip 5truept \hrule \vskip 2truept \hrule
                   \vskip 5truept} }


\footline={\ifnum\pageno>0 \tenrm \hss \folio \hss \fi }

\def\today
 {\count10=\year\advance\count10 by -1900 \number\day--\ifcase
  \month \or Jan\or Feb\or Mar\or Apr\or May\or Jun\or
             Jul\or Aug\or Sep\or Oct\or Nov\or Dec\fi--\number\count10}

\def\hour{\count10=\time\count11=\count10
\divide\count10 by 60 \count12=\count10
\multiply\count12 by 60 \advance\count11 by -\count12\count12=0
\number\count10 :\ifnum\count11 < 10 \number\count12\fi\number\count11}

\def\draft{
   \baselineskip=20pt
   \def\makeheadline{\vbox to 10pt{\vskip-22.5pt
   \line{\vbox to 8.5pt{}\the\headline}\vss}\nointerlineskip}
   \headline={\hfill \seven {\bcp Draft version}: today is \today\ at \hour
              \hfill}
          }

%
%

%
\catcode`@=11
%
%
\def\b@lank{ }

\newif\if@simboli
\newif\if@riferimenti

\newwrite\file@simboli
\def\simboli{
    \immediate\write16{ !!! Genera il file \jobname.SMB }
    \@simbolitrue\immediate\openout\file@simboli=\jobname.smb}

\newwrite\file@ausiliario
\def\riferimentifuturi{
    \immediate\write16{ !!! Genera il file \jobname.AUX }
    \@riferimentitrue\openin1 \jobname.aux
    \ifeof1\relax\else\closein1\relax\input\jobname.aux\fi
    \immediate\openout\file@ausiliario=\jobname.aux}

\newcount\eq@num\global\eq@num=0
\newcount\sect@num\global\sect@num=0

\newif\if@ndoppia
\def\numerazionedoppia{\@ndoppiatrue\gdef\la@sezionecorrente{\the\sect@num}}

\def\se@indefinito#1{\expandafter\ifx\csname#1\endcsname\relax}
\def\spo@glia#1>{} 

\newif\if@primasezione
\@primasezionetrue

\def\s@ection#1\par{\immediate
    \write16{#1}\if@primasezione\global\@primasezionefalse\else\goodbreak
    \vskip\spaziosoprasez\fi\noindent
    {\bf#1}\nobreak\vskip\spaziosottosez\nobreak\noindent}
%

\def\sezpreset#1{\global\sect@num=#1
    \immediate\write16{ !!! sez-preset = #1 }   }

\def\spaziosoprasez{50pt plus 60pt}
\def\spaziosottosez{15pt}

\def\sref#1{\se@indefinito{@s@#1}\immediate\write16{ ??? \string\sref{#1}
    non definita !!!}
    \expandafter\xdef\csname @s@#1\endcsname{??}\fi\csname @s@#1\endcsname}

\def\autosez#1#2\par{
    \global\advance\sect@num by 1\if@ndoppia\global\eq@num=0\fi
    \xdef\la@sezionecorrente{\the\sect@num}
    \def\usa@getta{1}\se@indefinito{@s@#1}\def\usa@getta{2}\fi
    \expandafter\ifx\csname @s@#1\endcsname\la@sezionecorrente\def
    \usa@getta{2}\fi
    \ifodd\usa@getta\immediate\write16
      { ??? possibili riferimenti errati a \string\sref{#1} !!!}\fi
    \expandafter\xdef\csname @s@#1\endcsname{\la@sezionecorrente}
    \immediate\write16{\la@sezionecorrente. #2}
    \if@simboli
      \immediate\write\file@simboli{ }\immediate\write\file@simboli{ }
      \immediate\write\file@simboli{  Sezione
                                  \la@sezionecorrente :   sref.   #1}
      \immediate\write\file@simboli{ } \fi
    \if@riferimenti
      \immediate\write\file@ausiliario{\string\expandafter\string\edef
      \string\csname\b@lank @s@#1\string\endcsname{\la@sezionecorrente}}\fi
    \goodbreak\vskip 48pt plus 60pt
\centerline{\lltitle #2}                     
\par\nobreak\vskip 15pt \nobreak\noindent}

\def\semiautosez#1#2\par{
    \gdef\la@sezionecorrente{#1}\if@ndoppia\global\eq@num=0\fi
    \if@simboli
      \immediate\write\file@simboli{ }\immediate\write\file@simboli{ }
      \immediate\write\file@simboli{  Sezione ** : sref.
          \expandafter\spo@glia\meaning\la@sezionecorrente}
      \immediate\write\file@simboli{ }\fi
\noindent\lltitle \s@ection#2 \par}


\def\eqpreset#1{\global\eq@num=#1
     \immediate\write16{ !!! eq-preset = #1 }     }

\def\eqref#1{\se@indefinito{@eq@#1}
    \immediate\write16{ ??? \string\eqref{#1} non definita !!!}
    \expandafter\xdef\csname @eq@#1\endcsname{??}
    \fi\csname @eq@#1\endcsname}

\def\eqlabel#1{\global\advance\eq@num by 1
    \if@ndoppia\xdef\il@numero{\la@sezionecorrente.\the\eq@num}
       \else\xdef\il@numero{\the\eq@num}\fi
    \def\usa@getta{1}\se@indefinito{@eq@#1}\def\usa@getta{2}\fi
    \expandafter\ifx\csname @eq@#1\endcsname\il@numero\def\usa@getta{2}\fi
    \ifodd\usa@getta\immediate\write16
       { ??? possibili riferimenti errati a \string\eqref{#1} !!!}\fi
    \expandafter\xdef\csname @eq@#1\endcsname{\il@numero}
    \if@ndoppia
       \def\usa@getta{\expandafter\spo@glia\meaning
       \la@sezionecorrente.\the\eq@num}
       \else\def\usa@getta{\the\eq@num}\fi
    \if@simboli
       \immediate\write\file@simboli{  Equazione
            \usa@getta :  eqref.   #1}\fi
    \if@riferimenti
       \immediate\write\file@ausiliario{\string\expandafter\string\edef
       \string\csname\b@lank @eq@#1\string\endcsname{\usa@getta}}\fi}

\def\autoeqno#1{\eqlabel{#1}\eqno(\csname @eq@#1\endcsname)}
\def\autoleqno#1{\eqlabel{#1}\leqno(\csname @eq@#1\endcsname)}
\def\eqrefp#1{(\eqref{#1})}


\def\eq{\autoeqno}
\def\req{\eqrefp}
\def\chap{\autosez}        



\newcount\cit@num\global\cit@num=0

\newwrite\file@bibliografia
\newif\if@bibliografia
\@bibliografiafalse

\def\lp@cite{[}
\def\rp@cite{]}
\def\trap@cite#1{\lp@cite #1\rp@cite}
\def\lp@bibl{[}
\def\rp@bibl{]}
\def\trap@bibl#1{\lp@bibl #1\rp@bibl}

\def\refe@renza#1{\if@bibliografia\immediate        
    \write\file@bibliografia{
    \string\item{\trap@bibl{\cref{#1}}}\string
    \bibl@ref{#1}\string\bibl@skip}\fi}

\def\ref@ridefinita#1{\if@bibliografia\immediate\write\file@bibliografia{
    \string\item{?? \trap@bibl{\cref{#1}}} ??? tentativo di ridefinire la
      citazione #1 !!! \string\bibl@skip}\fi}

\def\bibl@ref#1{\se@indefinito{@ref@#1}\immediate
    \write16{ ??? biblitem #1 indefinito !!!}\expandafter\xdef
    \csname @ref@#1\endcsname{ ??}\fi\csname @ref@#1\endcsname}

\def\c@label#1{\global\advance\cit@num by 1\xdef            
   \la@citazione{\the\cit@num}\expandafter
   \xdef\csname @c@#1\endcsname{\la@citazione}}

\def\bibl@skip{\vskip +4truept}


\def\stileincite#1#2{\global\def\lp@cite{#1}\global   
    \def\rp@cite{#2}}                                 
\def\stileinbibl#1#2{\global\def\lp@bibl{#1}\global   
    \def\rp@bibl{#2}}                                 

\def\citpreset#1{\global\cit@num=#1
    \immediate\write16{ !!! cit-preset = #1 }    }

\def\autobibliografia{\global\@bibliografiatrue\immediate
    \write16{ !!! Genera il file \jobname.BIB}\immediate
    \openout\file@bibliografia=\jobname.bib}

\def\cref#1{\se@indefinito                  
   {@c@#1}\c@label{#1}\refe@renza{#1}\fi\csname @c@#1\endcsname}

\def\cite#1{\trap@cite{\cref{#1}}}                  
\def\ccite#1#2{\trap@cite{\cref{#1},\cref{#2}}}     
\def\ncite#1#2{\trap@cite{\cref{#1}--\cref{#2}}}    
\def\upcite#1{$^{\,\trap@cite{\cref{#1}}}$}               
\def\upccite#1#2{$^{\,\trap@cite{\cref{#1},\cref{#2}}}$}  
\def\upncite#1#2{$^{\,\trap@cite{\cref{#1}-\cref{#2}}}$}  

\def\clabel#1{\se@indefinito{@c@#1}\c@label           
    {#1}\refe@renza{#1}\else\c@label{#1}\ref@ridefinita{#1}\fi}

\def\biblskip#1{\def\bibl@skip{\vskip #1}}           

\def\insertbibliografia{\if@bibliografia             
    \immediate\write\file@bibliografia{ }
    \immediate\closeout\file@bibliografia
    \catcode`@=11\input\jobname.bib\catcode`@=12\fi}


\def\commento#1{\relax}
\def\biblitem#1#2\par{\expandafter\xdef\csname @ref@#1\endcsname{#2}}


\catcode`@=12


\magnification=1200
\topskip 20pt
\def\interlinea{\baselineskip=16pt}
\def\standardpage{\vsize=20.7truecm\voffset=+1.truecm
                  \hsize=15.truecm\hoffset=+10truemm
                  \parindent=1.2truecm}

\tolerance 100000
\biblskip{+8truept}                        
\def\hbup{\hfill\break\baselineskip 16pt}  


\global\newcount\notenumber \global\notenumber=0
\def\note #1 {\global\advance\notenumber by1 \baselineskip 10pt
              \footnote{$^{\the\notenumber}$}{\nine #1} \interlinea}



\font\text=cmr10

\font\it=cmti10

\font\title=cmbx10 scaled \magstep3      
\font\ltitle=cmbx12 scaled \magstep1
\font\lltitle=cmbx12 scaled \magstep1
\font\subtitle=cmbx10 scaled \magstep1
\font\abs=cmti10 scaled \magstep1        

\font\seven=cmr7                         
\font\nine=cmr9                         
\font\bcp=cmbx7








\def\gtrsim{\ \rlap{\raise 2pt \hbox{$>$}}{\lower 2pt \hbox{$\sim$}}\ }
\def\lesssim{\ \rlap{\raise 2pt \hbox{$<$}}{\lower 2pt \hbox{$\sim$}}\ }


\def\mn{\medskip\noindent}
\def\bs{\bigskip}
\def\hb{\hfil\break}
\def\no{\noindent}

\def\o{\over}



\def\ea{{\it et.al.}}
\def\ib{{\it ibid.\ }}

\def\npb#1{Nucl. Phys. {\bf B#1},}
\def\plb#1{Phys. Lett. {\bf B#1},}
\def\prd#1{Phys. Rev. {\bf D#1},}
\def\prl#1{Phys. Rev. Lett. {\bf #1},}
\def\ncim#1{Nuo. Cim. {\bf #1},}
\def\zpc#1{Z. Phys. {\bf C#1},}
\def\prep#1{Phys. Rep. {\bf #1},}

\def\ijmpa#1{Int. Jour. Mod. Phys. {\bf A#1},}


\stileincite{}{}     
\numerazionedoppia   

\interlinea
\standardpage
\text                


\def\o{\over}


\def\G{{\cal G_{\rm SM}}}
\def\E{{\rm E}_6}


\def\pr{\prime}





\def\ie{{\it i.e.\ }}
\def\eg{{\it e.g.\ }}

\def\ra{\rangle}
\def\la{\langle}

\def\o{\over}


\def\nue{{$\nu_e\,$}}
\def\num{{$\nu_\mu\,$}}
\def\nut{{$\nu_\tau\,$}}

\def\E#1{{$E_{#1}\,$}}


\def\lu{{H^cQu^c}}
\def\ld{{HQd^c}}
\def\lt{{HLe^c}}
\def\lq{{S^chh^c}}
\def\lc{{hu^ce^c}}
\def\ls{{LQh^c}}
\def\lst{{\nu^chd^c}}
\def\lo{{hQQ}}
\def\ln{{h^cu^cd^c}}
\def\ldi{{H^cL\nu^c}}
\def\luu{{H^cHS^c}}


\def\l#1#2#3#4{\lambda^{^{\scriptstyle (#1)}}_{\scriptscriptstyle #2#3#4}\>}
\def\lw#1{\lambda^{^{\scriptstyle (#1)}}\>}

\def\G{{\cal G_{\rm SM}}}
\def\E{{\rm E}_6}

\def\pr{\prime}





\autobibliografia
\pageno=0
\vsize=23.8truecm
\hsize=15.7truecm
\voffset=-1.truecm
\hoffset=+6truemm
\baselineskip 12pt
\rightline{UM-TH 93--09}\par\noindent
\rightline{hep-ph/9304266}\par\noindent

\bs\bs\bs
\centerline{\ltitle Unconventional superstring derived E$_{\bf 6}$ models}
\bs
\centerline{\ltitle and neutrino phenomenology.}
\medskip
\medskip
\bs\bs                                   
                 \centerline{Enrico Nardi}
\bs
\centerline{\it Randall Laboratory of Physics, University of Michigan,
           Ann Arbor, MI 48109--1120}
\vskip 1truecm                             

\medskip
\centerline  { \bf {\abs Abstract}}      
\bs

\noindent
Conventional superstring derived $\E$ models can accommodate small
neutrino masses if  a discrete symmetry is imposed which forbids tree
level Dirac neutrino masses but allows for radiative mass generation.
Since the only possible symmetries of this kind are known to be
generation dependent, we explore the possibility that the three sets
of light states in each generation do not have the same assignments
with respect to the {\bf 27} of $E_6$, leading to non universal gauge
interactions under the additional $U(1)^\pr$ factors for the known
fermions. We argue that models realising such a scenario are viable,
with their structure being constrained mainly by the requirement of
the absence of flavor changing neutral currents in the Higgs sector.
Moreover, in contrast to the standard case, rank 6 models are not
disfavoured with respect to rank 5. By requiring the number of light
neutral states to be minimal, these models have an almost unique
pattern of neutrino masses and mixings. We construct a model based on
the unconventional assignment scenario in which (with a natural choice
of the parameters) $m_{\nu_\tau}\sim O(10)\,$eV is generated at one
loop, $m_{\nu_\mu}$ is generated at two loops and lies in a range
interesting for the solar neutrino problem, and ${\nu_e}$ remains
massless. In addition, since baryon and lepton number are conserved,
there is no proton decay in the model. In order to illustrate the
non-standard phenomenology implied by our scheme we also discuss a
second scenario in which an attempt for solving the solar neutrino
puzzle with matter enhanced oscillations and practically massless
neutrinos can be formulated, and in which peculiar effects for the
$\nu_\mu \rightarrow \nu_\tau$ conversion of the upward-going
atmospheric neutrinos could arise as well.
\bs
\noindent
PACS number(s): 12.10.Dm,13.15.Jr,12.15.-y,14.60.Gh
\vfill
\noindent
--------------------------------------------\phantom{-} \hb
\leftline{E-mail: nardi@umiphys.bitnet}
\bigskip
\leftline{UM-TH 93--09}
                   \medskip
\centerline{March 1993}

\eject

\standardpage                            
\interlinea                              

\null
\chap{Int} I. Introduction

It is generally believed that neutrinos possess very small but
non-vanishing masses.
While there is no fundamental reason for the neutrinos to be exactly
massless, small $\nu$ masses are needed in any particle physics
explanation of the solar neutrino problem, and at the same time they
imply several interesting phenomenological consequences.
A  very attractive way of generating naturally small neutrino masses
is through the use of the see-saw mechanism [\cite{see-saw}]. In $\E$
supersymmetric Grand Unified Theories (GUTs) [\cite{rizzo-e6}], as
derived from superstring theories, the see-saw mechanism cannot be
easily implemented since the Higgs representation necessary to
generate a large Majorana mass for the right-handed neutrinos is
absent.
However, even in the absence of Majorana terms, small masses can be
generated through radiative corrections in models in which at the
lowest order $m_\nu=0$. As was pointed out by Campbell et. al
[\cite{ellis-e6}] and Masiero et al. [\cite{MNS}], $\E$ GUTs do offer
the possibility of implementing this second mechanism.

The fermion content of models based on $\E$ is enlarged with respect
to the Standard Model (SM). In fact two additional lepton
$SU(2)$-doublets, two $SU(2)$-singlet neutral states and two
color-triplet $SU(2)$-singlet $d$-type quarks are present in
the fundamental representation of the group.
In order to forbid neutrino masses at the tree level an appropriate
discrete symmetry has to be imposed on the superpotential of the
model.  Branco and
Geng (BG) [\cite{branco}] have shown that no generation-blind symmetry
exists that forbids non vanishing neutrino masses at the tree level,
and at the same time allows for the radiative generation of the masses
at one loop. As a result, in order to implement this mechanism a
symmetry that does not act in the same way on the three generations is
needed.
In Section II we will briefly review the main fetures of the
conventional $\E$ models, and we will establish the notations.


Our work stems from the observation that once we chose
to build a model based on a symmetry that does distinguish among the
different generations, there is no reason in principle to expect that
this symmetry will result in a set of
{\it light} fermions  ({\ie the known states) that will
exactly replicate throughout the three generations.
To state this idea more clearly, we wish to suggest the possibility
that what we call ``\nut"
is actually assigned to an  $SU(2)$ doublet which has a different
embedding in $\E$ with respect to the doublet that contains
what we call ``\nue". As a consequence the two neutrinos
will have different $\E$ gauge interactions.
More drastically we can envisage the possibility
that the gauge interactions of
the $d$-quarks and leptons of one
family (say the third one) are different from those of the
corresponding states of the other two generations.
Obviously experimentally  we know that the $SU(2)\times U(1)$
interactions of the fermions do respect universality
with a high degree of precision,
however, in the class of models that we want to investigate one or two
additional $U(1)^\pr$ abelian factors are always present, implying
additional massive neutral gauge bosons possibly at energies $O$(TeV)
or less. The possibility that the $U(1)^\pr$ interactions of the
known fermions could violate universality then is indeed still
phenomenologically viable.

In Section III we will develop a scenario that realises this idea.
Starting from the assumption of Unconventional Assignments
(UA) for the light neutrino of the third generation, we will show
that the need for UA is reflected in the $d$-quark sector as well,
thus leading to a third generation of light fermions which is not
a replica of the first two.
In Section IV we will concentrate on the neutrino phenomenology,
and we will describe the pattern of masses and mixings that
is predicted by our scheme.
We believe that the unconventional scenario that we are going to
analyse here could be interesting in itself, since it is not a priori
obvious that models in which the `low' energy gauge interactions of the known
fermions are not universal can be consistently constructed. However, it
turns out that beyond being viable, these models also lead to an
interesting phenomenology, expecially in the neutrino sector, and
as well imply some rather unusual consequences.
In order to illustrate this,
at the end of Section IV we will discuss a particular model in which
a few peculiar effects in
the propagation of the neutrinos through matter could arise.
We will formulate
an attempt to find a Mikheyev-Smirnov-Wolfenstein
(MSW)-like solution to the solar neutrino puzzle [\cite{MSW}]
with `almost massless' neutrinos ({\ie $m_\nu \ll 10^{-3}$ eV}).
We will also address the issue that
these non-standard effects
could lead to a suppression of the ``\num''--``\nut'' oscillations
for the high energy upward-going atmospheric neutrinos.
Finally in Section V we will summarize our results and draw the conclusions.


\chap{conv} II. Conventional E$_{\bf 6}$ models.

In $\E$ GUTs, matter fields belong to the fundamental {\bf 27}
representation of the group. $\E$ contains $SO(10)\times U_\psi(1)$ as
a maximal subalgebra, and the {\bf 27} branches to ${\bf 1} + {\bf 10}
+ {\bf 16}$ of $SO(10)$.
In turn $SU(10)$ contains $SU(5)\times U_\chi(1)$.


\def \tabrul2     {\noalign{\vskip 5truept \hrule \vskip 2truept \hrule
                   \vskip 5truept} }
\def\om{\omit}

\baselineskip=12pt
\midinsert
{
\bs
$$
\vbox{\hsize= 13.3truecm
\baselineskip=14pt
\noindent TABLE I.
{\nine
$SO(10)$, $U_\psi(1)$, $SU(5)$ and $U_\chi(1)$ assignments
for the left--handed fermions of the {\bf 27} fundamental
representation of ${\rm E}_6$. The $SU(2)$ doublets
$H^c$, $H$, $L$ and $Q$ are
explicitly written in components.
The Abelian charges $q_\psi$ and $q_\chi$ can be derived
from the quantum numbers listed in the square
brackets by dividing by $c_\psi = 6\sqrt{2/5}$
and $c_\chi = 6\sqrt{2/3}$ respectively.
The charges
are normalized to the hypercharge axis according to:
$\sum_{f=1}^{27}(q_{\psi,\chi}^f)^2
= \sum_{f=1}^{27}({1 \over 2}Y^f)^2 = 5$.
}
\vskip -.8truecm}
$$

$$
\vbox{
\hsize= 13.35truecm
\offinterlineskip
\halign {
\vrule#&\strut#&\vrule#&\strut#&\vrule#&\strut\quad#&\vrule#&\strut#&
\vrule#&\strut#&\vrule#&\strut#&\vrule#&\strut\quad#&\vrule#&\strut#&
\vrule#&\strut#&\vrule#&\strut#&\vrule#&\strut#&\vrule#&\strut#&\vrule#\cr
\noalign{\hrule}
height6pt & \om && \om && \om&\om&\om && \om&\om&\om && \om &&
\om&\om&\om && \om&\om&\om&\om&\om &\cr
&\om && $\ S^c\> $ && $\> {{E^c}\choose{N^c}}\ $&\om&$\ h\ $ && $\
{N\choose{E}}\ $&\om&$\ h^c\ \ $ && $\> \nu^c\ $ && $\ \>
{\nu\choose{e}}$&\om&$\ d^c\ \>$ && $\ \> e^c\ \>$&\om&$\ u^c\
\>$&\om&$\ {u\choose d}\ \>$ &\cr
height6pt&\om&&\om&&\om&\om&\om&&\om&\om&\om&&\om&&
\om&\om&\om&&\om&\om&\om&\om&\om&\cr
\noalign{\hrule}
height6pt&\om&&\om&&\om&\om&\om&\om&\om&\om&\om&&
\om&\om&\om&\om&\om&\om&\om&\om&\om&\om &\om&\cr
& $\ SO(10)\> (c_\psi q_\psi)\ $ && $\ {\bf 1}\>(4)\ $ &&   
\multispan7 \hfil ${\bf 10}\> (-2)$\hfil          
&&                                                
\multispan{11} \hfil ${\bf 16}\> (1)$ \hfil       
&\cr
height6pt&\om&&\om&&
\om&\om&\om&\om&\om&\om&\om&&
\om&\om&\om&\om&\om&\om&\om&\om&\om&\om&\om&\cr
\noalign{\hrule}
height6pt&\om&&\om&&\om&\om&\om&&\om&\om&\om&&
\om&&\om&\om&\om&&\om&\om&\om&\om&\om&\cr
& $\ SU(5)\ (c_\chi q_\chi)\  $ && $\ {\bf 1}\>(0)\ $ && \multispan3
\hfil ${\bf 5}\>(2)$ \hfil           
&& \multispan3 \hfil ${\bf
\bar 5}\>(-2)$\hfil                  
&& \hfil ${\bf 1}\>(-5)\ $ \hfil
&&                                   
\multispan3 \hfil ${\bf \bar 5}\>(3)$\hfil        
&&                                                
\multispan5
\hfil ${\bf 10}\> (-1)$\hfil                      
&\cr
height6pt & \om && \om && \om&\om&\om && \om&\om&\om && \om &&
\om&\om&\om && \om&\om&\om&\om&\om &\cr
\noalign{\hrule}}}
$$
\bs
}
\endinsert
\interlinea

The $SO(10)$,  $SU(5)$, $U_\psi(1)$ and $U_\chi(1)$ assignments for
the states in the {\bf 27} are listed in Table I. Usually the known
particles of the three generations are assigned to the {\bf 16} of
$SO(10)$ that also contains a right handed neutrino :
$$
[{\bf 16}]_i = \big[
Q\equiv { 
{u\choose d}}, \,
u^c, \,
e^c, \,
d^c, \,
L\equiv { 
{\nu \choose e}}, \,
\nu^c\big]_i \qquad \qquad i=1,2,3
\eq{2.1}
$$
\mn
while the {\bf 10} and the {\bf 1} of $SO(10)$
contain the following new fields
$$
\eqalign{
[{\bf 10}]_i &= \big[
H^c\equiv {  
{E^c\choose N^c}}, \,
h, \,
H\equiv {  
{N\choose E}}, \,
h^c\big]_i                                          \cr
[\>{\bf 1}\>]_i&= [S^c]_i   \qquad \qquad \qquad \qquad
\hskip 2.3truecm i=1,2,3. \cr}
\eq{2.2}
$$
\mn
As it is clear from Table I
there is an ambiguity in assigning the known states to the {\bf 27},
since under the SM gauge group $\G\equiv
SU(3)_c\times SU(2)_L\times U(1)_Y$ the ${\bf \bar 5}_{({\bf 10})}$ in
the {\bf 10} of $SO(10)$ has the same field content as the ${\bf
\bar 5}_{({\bf 16})}$ in the {\bf 16}. The same ambiguity is
also present for the two $\G$ singlets, namely
 ${\bf 1}_{({\bf 1})}$ and ${\bf 1}_{({\bf 16})}$.

Since $\E$ is rank 6 as many as two additional neutral gauge bosons
can be present, corresponding for example to some linear combinations
of the $U_\chi(1)$ and $U_\psi(1)$ generators.
The fermion interactions with
these gauge bosons will depend on the specific assignments. The
two additional neutral gauge bosons are usually parametrized as
$$
\eqalign{
&Z_\beta^\pr= Z_\psi \sin\beta + Z_\chi \cos\beta  \cr
&Z_\beta^{\pr\pr}= Z_\psi \cos\beta - Z_\chi \sin\beta, \cr}
\eq{2.3}
$$
\mn
and in the following we will often collectively
refer to them as $Z_\beta$ bosons.
In the presence of at least one `light' $Z_\beta$
($M_{\beta}\lesssim 1-2\,$TeV),
different assignments will lead to a different phenomenology
that could be tested \eg at LHC and SSC energies or possibly even at
LEP II.
In contrast it is clear that in the limit $M_{\beta}\rightarrow \infty$
the choice of the assignment is irrelevant as far as
we are only concerned with the gauge interactions.  However, as we
will show,  even in
this limit the requirement of $U_\beta(1)$ gauge invariance for the
superpotential, together with the phenomenological constraints on the
absence of FCNC in the Higgs sector, will have far reaching consequences
for determining the structure of the viable models.




The most general renormalizable superpotential
arising from the coupling of
the three {\bf 27}'s in Table I and invariant under the
SM gauge group is [\cite{e6-super}]
$$
W=W_1+W_2+W_3+W_4
$$
\noindent where
$$
\eqalign{
&W_1=\lw1\lu+\lw2\ld+\lw3\lt+\lw4\lq \cr
&W_2=\lw5\lc+\lw6\ls+\lw7\lst        \cr
&W_3=\lw8\lo+\lw9\ln                 \cr
&W_4=\lw{10}\ldi+\lw{11}\luu.          \cr
}
\eq{2.4}
$$
\mn
The Yukawa couplings in \req{2.4} are three index tensors in
generation space, \eg
$\lw1\lu \equiv \l1ijk H^c_iQ_ju^c_k\>$
with $i,j,k = 1,2,3$ generation indices,
and  in general they are not constrained by the
$\E$ Clebsch-Gordan relations [\cite{witten-yuk}].
The presence of $W_4$ would produce tree level
Dirac masses for all the neutral states in the model.
In particular the $\nu$'s  
would acquire a Dirac mass
$m_\nu = \lw{10}\langle \tilde{ N^c}\rangle$.
An unnatural tuning of the $\lw{10}$ Yukawa
couplings is then required to make these masses small.
If $\lw{10}$ were absent, then
$\nu$ and $\nu^c$  
would be massless at tree level.
Furthermore, if at the same time the couplings
$\lw6$ and $\lw7$ in $W_2$ were non-vanishing,
naturally small Dirac masses would be produced
at the one-loop level through diagrams like the one depicted in Fig. 1
[\cite{ellis-e6},\cite{MNS}}].
However a problem arises due to the fact that
the simultaneous presence of $W_2$
$(\supset \lw6,\>\lw7)$ and $W_3$
induces fast proton decay.
The vanishing of $W_3$ can cure this problem still allowing
for radiative neutrino masses.
The conclusion is that in the conventional schemes the vanishing
of $W_3$ and $\lw{10}$ together with non-vanishing
$\lw6$ and $\lw7$ couplings is required in order to have an
interesting neutrino phenomenology and not to conflict with the limits
on the proton lifetime. As was discussed in detailed by
BG [\cite{branco}] the correct pattern of vanishing Yukawa couplings
leading to small \nue, \num and \nut masses
can be realised only by means
of a generation-dependent discrete symmetry, \ie a symmetry under
which the fields transform with a generation dependent phase
$\psi_j \rightarrow e^{i\alpha_{\psi_j}} \psi_j$
where $j$ is a generation index.

\midinsert
{
\epsfxsize=13.3truecm           
\epsffile{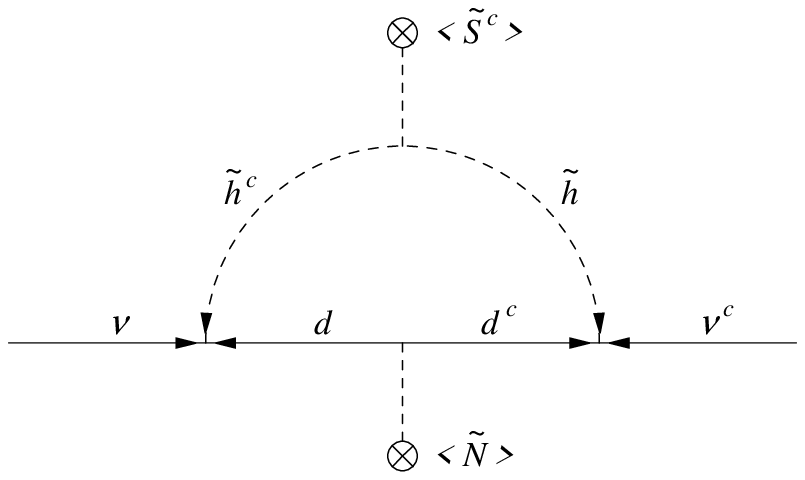}
$$
\vbox{\hsize=13.3truecm
\baselineskip=12pt
\noindent Fig. 1:
{\nine
A typical diagram contributing to the neutrino Dirac masses
at the one-loop level.}
}
$$
\bs
}
\endinsert
\interlinea

\chap{flip} III. The unconventional assignment scenario.

Once we give up the assumption that
from the point of view of the symmetry transformation properties
the three generations
are exact replicas one of the other,
we may also abandon the assumption that
the known particles of the three  generations
should be identified with the same states in the three
different {\bf 27}'s.
We will now explore the possibility
of constructing a consistent model in which
the assignments of the known fermions to the {\bf 27}
are different for the different generations.
Models of these kind turn out to be phenomenologically viable, and
clearly
they imply a few unusual phenomenological consequences. For example,
the known fermions will
have generation dependent
Neutral Current (NC) gauge interactions induced by
$Z_\beta$ exchange, due to the difference in the $U_\psi(1)$ and
$U_\chi(1)$ charges.

\bigskip
\no{\subtitle The assignments for the leptons. }
\medskip
\nobreak
As a starting point for investigating $\E$ models with
UA we will assume that what we call ``$\nu_\tau$"
is in fact the $N_3$ weak doublet neutral state belonging to the
${\bf \bar 5}_{{\bf 10}}$, while
\nue and \num are still assigned as usual to the
${\bf \bar 5}_{{\bf 16}}$.
We will henceforth use quotation marks to denote the known states with
their  conventional labels, since they might not correspond to the
entries in Table I. Labels not enclosed within quotation marks will
always refer to the fields listed as in the  Table.
We will also keep the same assignments as well as the same transformation
properties under the discrete symmetry group
for the {\it known} states of the first and second
generations,
(other assignments leading to different models are trivially obtained
by interchanging the generation labels  \eg
$1 \leftrightarrow 3$).
Accordingly, when we
refer to the first generation it is understood that the same
applies to the second generation as well.
Latin indices $i$, $j$\dots \ run
from 1 to 3, while the greek indices $\alpha$, $\beta$ \dots
$=1,2$ will refer only to the first two generations.

With the notation given in \req{2.1} and \req{2.2}, and
referring to the {\bf 10} and {\bf 16}
of $SO(10)$, our starting assumption for the assignments of the
three $SU(2)$ doublet light neutrinos reads:
$$
\eqalign{
&``\nu_\alpha" \in L_\alpha \in {\bf 16} \qquad\qquad \alpha=1,2 \cr
&``\nu_\tau"  \in H_3 \in {\bf 10}. \cr            }
\eq{2.5}
$$
\mn
In order to realise this scenario we first have to require
that the tree level masses for
$\nu_\alpha$ and $N_3$ vanish. This can be achieved by setting
$$
\l{10}{\la i\ra}\alpha j (H^c_i L_\alpha \nu^c_j) = 0
\qquad\qquad
\l{11}{\la i\ra}3j       (H^c_i H_3      S^c_j) = 0.
\eq{2.6}
$$
\mn
For the sake of clarity we have enclosed inside
$\la$brackets$\ra$
the indices labeling the particular Vacuum Expectation Values
(VEVs) which are relevant
for the actual discussion.
{}From the LEP measurement of the number of weak-doublet neutrinos we
know that all the remaining $SU(2)$ doublet neutral states
$N_\alpha$, $\nu_3$ and $N^c_i$ must be heavy ($\gtrsim 50\,$GeV).
This in turn implies that the following terms must be non-vanishing :
$$
\l{10}i3{\la\beta\ra}      (H^c_i L_3     \nu^c_\beta) \neq 0
\qquad\qquad
\l{11}i\alpha{\la\beta\ra} (H^c_i H_\alpha  S^c_\beta) \neq 0.
\eq{2.7}
$$
If any two of the scalar components of the three neutral fields
in the trilinear terms $\lw{10}$ and $\lw{11}$
acquire a VEV, then the
VEV of the third scalar field must also be non-vanishing.
Therefore, almost all of the neutral scalars in $W_4$ (doublets and singlets)
will
eventually acquire a VEV. In particular it is not difficult to see
that in order to have all the $H^c_i$ heavy, none of their scalar
component can be prevented from eventually acquiring a VEV.
This is the reason why
we have forbidden all the couplings between the massless
neutrinos and the $H^c_i$
fields in \req{2.6}.
We note that at the same time the conditions \req{2.6}
allow for $\langle\tilde \nu_\alpha\rangle
=\langle\tilde N_3\rangle = 0$ which,
as we will discuss, is necessary if we want to
prevent spontaneous violation of lepton number.

Due to our choice of the light states and due to the vanishing
of the couplings in \req{2.6} the following VEVs can be generated:
$\langle\tilde{\nu^c_\alpha}\rangle$,
$\langle\tilde{S^c_\alpha}\rangle$,
$\langle\tilde{N_\alpha}\rangle$
$\langle\tilde{\nu_3} \rangle$ and
$\langle\tilde{N^c_i}\rangle$.
It was argued in Ref. [\cite{ellis-e6}] that
in the conventional $\E$ models it might be difficult to achieve
$\langle\tilde{\nu^c} \rangle\ne 0$ since the Yukawa couplings
$\lw{7}$ and/or $\lw{10}$
that are needed for driving $m^2_{\tilde{\nu^c}}$ negative through the
renormalization group, are constrained to be either vanishing or
too small to generate
this VEV. This implies that the
set of VEVs present in the conventional models does not allow for lowering
the rank of the gauge group by more than one, and since
the SM is rank 4 it is probably not possible to construct a dynamical
model based on rank 6.
In contrast we will see that in the present scheme
some of the $\lw{7}$ are not constrained to be particularly small,
and indeed some of the $\lw{10}$
couplings (those in \req{2.7}) are expected to be rather large.
We can conclude that rank 6 models are indeed viable
in our UA scenario since $\langle\tilde{\nu^c} \rangle\ne 0$
can be easily achieved.

Now, in order to allow for radiatively generated Dirac masses, we need
massless right-handed neutrinos as well.
For the sake of simplicity, we will require
a {\it minimum} number of light neutral
$SU(2)$ singlets.
In \req{2.7} we have already assumed that the couplings involving
$\nu^c_3$ and $S^c_3$ are forbidden, thus preventing
their fermionic component from acquiring a mass at tree level.

Another consequence of \req{2.7} regards
the charged lepton mass matrix. In fact it is clear that
the $e_\alpha$ fields and the left-handed $``\tau"$ lepton $E_3$ have to
acquire
their mass from the $\lw3$ Yukawa coupling, since the $\lw{10}$
and $\lw{11}$ couplings for these states are forbidden.
Then the ``$\tau$'' lepton mass term $m_\tau E_3e^c_3$ must be
generated from the $\tilde {L_3}$ scalar doublet,
while $m_e$ and $m_\mu$ are generated from the VEV of
one of the $\tilde{H_\alpha}$
Higgs multiplet (say $\tilde {H_2}$).
As a consequence of this it is true that
in general all the right-handed leptons $e^c_j$
will couple to both $\tilde {L_3}$ and $\tilde {H_2}$
through the
couplings $\l33{\la 3\ra}j$ and $\l3{\la2\ra}\beta j$.
It is well known that this situation can
give rise to dangerous Lepton Flavor Violating (LFV) couplings
between the fermions and the Higgs fields [\cite{weinberg-fc}]
since the rotation that
diagonalizes the lepton mass matrix does not diagonalize
the fermion couplings to the Higgs fields.
In this respect the couplings with $\tilde {H_1}$ are also  dangerous
since its neutral scalar component will eventually acquire a VEV as well.
In addition, non-zero mass terms connecting
$E_\alpha$-$e^c_j$ ($e_3$-$e^c_j$)
which can be generated by non-vanishing
$\l3\alpha {\la3\ra} j$ ($\l3{\la\alpha\ra} 3 j$)
couplings will induce an
isospin violating ($\Delta I=1/2$) light-heavy
mixing between the $e^c$ and the $E^c$ fields.
It is well known [\cite{ellis-e6},\cite{ll1}]
that a mixing of this kind can give rise to
tree level LFV processes mediated  by $Z_0$ exchange.
Therefore, in order
not to conflict with the tight limits on LFV processes such
as $\mu\rightarrow eee$, $\mu$--$e$ conversion in muonic atoms, etc\dots,
we have to require all the $\lw3$ couplings to be absent, with
the exception of $\l3{\la2\ra}\alpha\beta$ and
$\l33{\la3\ra}3$ which are needed
to generate masses for the light leptons.
Together with the conditions in \req{2.6}, this
additional requirement insures that
all the light-heavy lepton mixings are absent.
In addition the resulting mass matrix for the light lepton
turns out to be block diagonal, with
$$
[m_{\ell}]_{\alpha \beta}=\l3{\la2\ra}\alpha\beta \langle {\tilde
{N_2}}
\rangle
\qquad \qquad
[m_{\ell}]_{33}=\l33{\la3\ra}3 \langle{\tilde {\nu_3}}\rangle
\eq{2.8}
$$
We note that an important consequence of
the constraints just discussed is that
any possible mixing of the third generation neutrino
can only arise in the neutrino sector.

\bigskip
\no{\subtitle The assignments for the quarks.}  
\medskip
\nobreak
At this stage three $SU(2)$-doublet and two $SU(2)$-singlet
neutral states
are massless, namely $\nu_\alpha$, $N_3$ and $\nu^c_3$, $S^c_3$.
Dirac masses for these states
cannot be generated via loops of leptons, since this
would require some of the couplings in \req{2.6}, but they can indeed
be induced by loops involving quarks through a set of
diagrams that are the analogous to the one depicted in Fig. 1.

The relevant couplings for generating these diagrams
are $\lw2\ld$, $\lw4\lq$, $\lw6\ls$ and $\lw7\lst$.
Besides appearing as vertices for the external states, each of these
couplings will also provide a
mass insertion for the quarks running inside the loop.
We have recast these couplings into
the schematic partial diagrams $A$ and $B$ depicted in
Fig. 2. In these diagrams it is understood that
one of the two neutral field is external
while the other one corresponds to a VEV insertion.
We will label $A_1$ and $B_1$ the partial diagrams $A$ and $B$
in which the VEV insertions correspond respectively
to the $SU(2)$ singlets $\la\tilde{S^c_\alpha}\ra$ and
$\la\tilde{\nu^c_\alpha}\ra$,
and $A_2$ and $B_2$ those diagrams
in which the VEV insertion corresponds to the
doublets $\la\tilde{\nu_3}\ra$ and $\la\tilde{N_\alpha}\ra$.

\midinsert
{
\epsfxsize=13.3truecm 
\epsffile{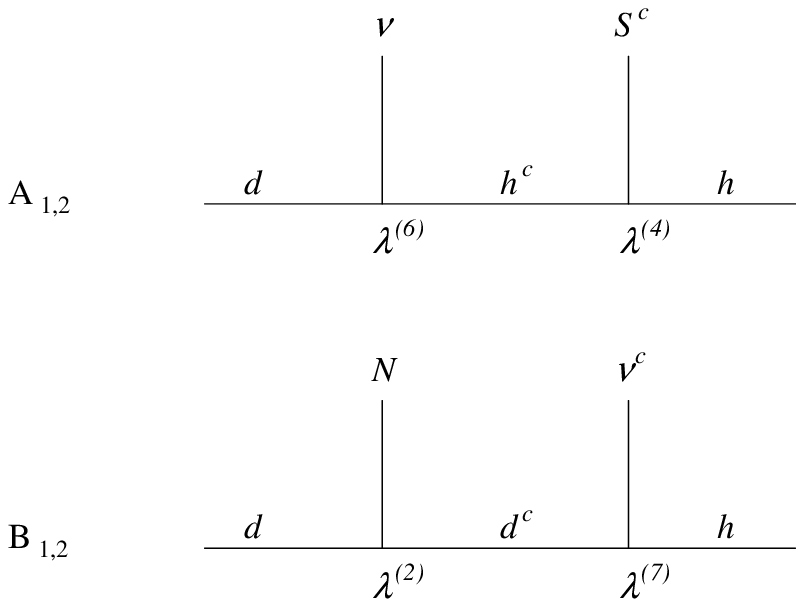}
$$
\vbox{\hsize= 13.3truecm
\baselineskip=12pt
\noindent Fig. 2:
{\nine
Schematic partial diagrams for generating one-loop mass entries in the
Dirac neutrino mass matrix. Both in $A$ and $B$ it is understood that
one of the two neutral field is external, while the other one
represents a VEV insertion that provides a mass for the quark running
inside the loop.
$A_1$ and $B_1$ correspond respectively to the diagrams where the
mass insertions are provided by the
VEVs $\la\tilde{S^c_\alpha}\ra$ and $\la\tilde{\nu^c_\alpha}\ra$
while $\nu_\alpha$ and respectively $N_3$ are external.
$A_2$ and $B_2$ correspond to insertions of the doublets VEVs
$\la\tilde{\nu_3}\ra$ and $\la\tilde{N_\alpha}\ra$ while
$S^c_3$ and $\nu^c_3$ are external.}
}
$$
\bs
}
\endinsert
\interlinea
\noindent
By gluing two partial diagrams together, we can generate
the following entries in the neutrino Dirac mass matrix:
$$
\pmatrix{\nu_\alpha & N_3 \cr}
\pmatrix{[A_1A_2] & [A_1B_2] \cr
         [B_1A_2] & [B_1B_2] \cr}
\pmatrix{S^c_3 \cr \nu^c_3 \cr}.
\eq{2.9}
$$
Obviously not all the couplings $\l mijk \>$ ($m=2,4,6,7$) are allowed,
and we will now proceed to select the couplings that must be forbidden.
In the first place we note that in order to generate a diagram that
will provide a radiative mass for $N_3$,
at least one of the couplings $\l23ij H_3 Q_i d^c_j$
in $B_1$ must be non-vanishing. However
$\lw2$ is precisely the Yukawa that, in the conventional models,
is needed to give mass to the down-type quarks, say
$[m_d]_{ij} = \l2{\la1\ra}ij \langle \tilde {N_1}\rangle d_i d^c_j$.
Then it is clear that
$N_3$ cannot couple to a pair of $d$-$d^c$ fields that acquire a mass
trough $\lw2$ otherwise it will necessarily have to transform
in the same way as $N_1\,$ does
under the discrete symmetry,  and $N_3$ will be forced to acquire
a large tree level mass through the same mechanism that makes $N_1$ heavy.
We conclude that the requirement of generating a one loop mass
for $N_3$ implies that some of the ``$d$" quarks cannot acquire
their mass through $\lw2$.
Assuming that all the ``$d$" quarks acquire a mass at the tree level,
implies that we need to flip the assignments
for some of the light right-handed $Q=-1/3$ fields as well.
More precisely if $d_{\underline i}d^c_{\underline j}$
are the fields coupled to $N_3$, the (light) $SU(2)$ doublet
field $d_{\underline i}$ has to acquire its mass through
the term $\l6{\la3\ra}{\underline i}k L_3 Q_{\underline i} h^c_k$
while the (heavy) $SU(2)$ singlet field
$d^c_{\underline j}$ will acquire its mass trough a singlet VEV
from the term $\l7{\la\alpha\ra} k{\underline j}
\nu^c_\alpha h_k d^c_{\underline j}$.
The same argument implies that we cannot flip
the assignments for {\it all} the ``$d$" quarks. In fact the need for
generating a radiative mass entry for $\nu_\alpha$ implies that some
$\l6\alpha i j L_\alpha Q_i h^c_j$ vertices in $A_1$
must be non vanishing.
At the same time the quark fields entering this  vertex
cannot acquire a $\lw6\la \tilde {\nu_3}\ra$ mass,
otherwise we could not prevent $\nu_\alpha$ from acquiring a large mass
like $\nu_3$.

All these requirements are satisfied for example
by the following assignments:
$$
\eqalign{
&``d^c_\alpha"\equiv h^c_\alpha \in {\bf 10} \qquad\qquad \alpha=1,2 \cr
&``d^c_3" \equiv d^c_3 \in {\bf 16}, \cr            }
\eq{2.10}
$$
meaning that
the massive states corresponding to the
known (light) $Q=-1/3$ quarks having $SU(2)$ chiral interactions
are $(d_\alpha\,h^c_\beta)$ and
$(d_3\,d^c_3)$ with their mass generated respectively
by $\langle \tilde {\nu_3}\rangle$
and  $\langle \tilde {N_1}\rangle$,
while  the heavy vectorlike $SU(2)$ singlets
$(h_i\,d^c_\beta)$ and $(h_i\,h^c_3)$
acquire a mass through the singlet VEVs
$\la \tilde{\nu^c_\alpha}\ra$
and $\la \tilde{S^c_\alpha}\ra $ ($\alpha=1$ and/or 2)
respectively.
In order to realise this scenario we are once again faced with
a problem of FCNC that must be confronted with
the tight experimental limits derived mainly from analyses
of the $K$ and $B$ systems. A large number
of Yukawa couplings must be forbidden in order
to avoid an excessive tuning for those parameters responsible
for the FCNC processes.
In order to avoid the
Higgs mediated FCNC in the light ``$d$" quarks sector,
which are a direct consequence of the
asymmetric assignments among the three families,
we must require $\l2{\la1\ra}\alpha3=\l6{\la3\ra}3\alpha=0$
and we must forbid all the couplings
between $(d_\alpha h^c_\beta)$ and $H_2$ as well.
Moreover, as in the charged lepton case,
the $\Delta I=1/2$ light-heavy mixing between the
$d_i$ and $h_j$ states will induce
$Z_0$ mediated FCNC [\cite{ellis-e6},\cite{ll1}].
In addition in the present case a new source of FCNC is represented
by the $\Delta I=0$ light-heavy mixings  among
$h^c_\alpha$, $d^c_3$ and $d^c_\beta$, $h^c_3$.
Unlike the $\Delta I=1/2$ case,
these mixings are not suppressed by any small doublet-VEV
to singlet-VEV ratio, and are then expected to be large [\cite{lfc}].
However, since they do not violate weak-isospin, no
FCNC processes can be induced by $Z_0$ exchange.
Nevertheless these mixings do still affect the $Z_\beta$ couplings,
and could indeed constitute an additional dangerous source of FCNC
in the presence of a $Z_\beta$ with mass below $\sim 1\,$ TeV
[\cite{lfc}].
Both these additional sources of  FCNC can be avoided by setting
$\l2{\langle \alpha\rangle} j\beta=\l6{\langle 3\rangle}j3=0$
and $\l4{\la\alpha\ra}j\beta=\l7{\la\alpha\ra}j3=0$.
In particular, we note that the second condition
is also needed for the sake of keeping a well defined meaning
to our UA, since in principle there is no reason
to expect that the $\lw4$ and $\lw7$
couplings generating the $h_j$-$h^c_\beta$ and
$h_j$-$h^c_3$ mixing mass terms should
be much smaller than the $\lw6$ and $\lw2$ Yukawa couplings
responsible, in our scheme,
for the ``$d$", ``$s$" and ``$b$" masses.

After all these conditions are implemented,
there are no light-heavy mixings in the whole quark sector.
The mass matrix for the light down-quarks reads
$$
\eqlabel{2.11}
\eqlabel{2.12}
\eqalignno{
&[m_{\cal D}]_{\alpha\beta}=
\l6{\la3\ra}\alpha\beta\langle \tilde {\nu_3}\rangle
&\req{2.11}  \cr
&[m_{\cal D}]_{33}=
\l2{\la1\ra}33\langle \tilde {N_1}\rangle.
&\req{2.12}   \cr
}
$$
The remaining $SU(2)$ singlets
$Q=-1/3$ quark states $h_i$, $d^c_\beta$ and $h^c_3$
are vectorlike, and acquire (large)
masses through VEVs of singlets
$$
\eqlabel{2.13}
\eqlabel{2.14}
\eqalignno{
&[M_{\cal D}]_{i\beta} =
\l7{\la\alpha\ra}i\beta \langle \tilde {\nu^c_\alpha}\rangle &\req{2.13}  \cr
&[M_{\cal D}]_{i3} =
\l4{\la\alpha\ra}i3\langle \tilde {S^c_\alpha}\rangle.   &\req{2.14}   \cr
}
$$
\mn
{}From \req{2.11} and \req{2.12}
we see that our starting assumption
about the flipped assignments for the ``$\tau$" neutrino
has had the far reaching consequence that the down-quark mass matrix
is block diagonal. Then all the mixing of the third family can be
generated only in the up-quark sector (this might as well suggest
a mechanism for explaining
the smallness of these mixings relative to the Cabibbo mixing).
As a result, in order to have a CKM matrix without zero entries,
the up-quark mass matrix must be truly 3$\times$3.
We note that
the $Q_3$ doublet cannot transform under the discrete symmetry
like the $Q_\alpha$ doublets, otherwise it would not be possible to
forbid the $d_\alpha$-$d^c_3$ mass terms and simultaneously
allow  for a non-zero $d_3$-$d^c_3$ mass.
At the same time in order to allow for
a 3$\times$3 up-quark mass matrix without zero entries
all  the $u^c_i$ fields
appearing in $\lw1$ have to
transform with the same phase.
Then since the bilinears $Q_\alpha u^c_j$ and $Q_3 u^c_j$
have to transform with an overall different discrete phase,
in order to construct trilinear invariants
they must be coupled to different $H^c$ Higgs fields.
As a conclusion, we see that
Higgs mediated FCNC cannot be completely avoided in the UA scenario.
However, in the scheme we are analysing here
they appear only in the up-quark sector.
Since there are no experimental data on FCNC involving the $t$ quark,
the only existing constraints are for the $u$-$c$ transitions.
The strongest bounds on these FCNC
come from the limits on $D^0$--${\bar {D^0}}$ oscillations that
receive contributions from
the $c{\bar u}\rightarrow {\bar c} u$ amplitude.
Other rare processes, as the rare decays $D\rightarrow \pi\pi, \> KK $
that could be induced by the
$c{\bar u}\rightarrow {\bar u} u$ amplitude, do not give
additional constraints.

A bound $\lw1 \sim \lambda^\pr_{uc} < 2\cdot 10^{-4}$ on the
off-diagonal $u$-$c$ coupling was obtained
in Ref. [\cite{ellis-e6}] from the limits on $D^0$--${\bar {D^0}}$
oscillations and assuming $M_{H^c} \simeq 100\,$GeV.
Since the $\lw1$ couplings are also responsible for
generating the up-quark mass ($\sim$ few MeV)
from VEVs $\sim 100$ GeV,  we indeed expect some of the $\lw1$
to be of order $10^{-4}$ or less.
We can conclude that the previous bound does not constitute a serious
constraint for the UA scheme, since it does not require a particular
tuning of the FCNC parameters. However
a definite prediction of the present model is the existence of an
amount of FCNC in the up-quark sector which is larger than in the SM.

Up to this point we have analysed the requirements which the
superpotential \req{2.4} must satisfy in order to realise
the UA scenario and produce an acceptable and possibly
interesting phenomenology. We
have carried out a general discussion without referring to
any particular discrete symmetry.
However, since  we have required the set of Yukawa couplings
in the superpotential \req{2.4} to satisfy
a rather large number of constraints, it could well be that
one particular transformation for one field, which is  needed in order to
forbid a dangerous coupling, at the same time implies
the vanishing of another coupling that we want to be non-zero.
A general proof that
our set of constraints is self-consistent
would be lengthy and cumbersome.
However it is enough to prove that there is at least one set of discrete
transformations for the fields that satisfies all our constraints,
and this will automatically
insure that our set of constraints is not
self-contradictory.

We have found that our scheme can be implemented by imposing on the
superpotential \req{2.4} a simple $Z_2\times Z_3$ symmetry. The
transformation properties of the fields in the three {\bf 27}'s are
listed in Table II. Beyond satisfying to all our requirements
it is easily seen that this
symmetry also insures that there is no
fast proton decay, since it forbids the terms
$\lw8$ and $\lw9$ all together (note that
in spite of the UA
these still represent the dangerous terms, being
both invariant under the exchange $d^c \leftrightarrow h^c$).

In more generality it can be shown
that proton stability is just a consequence of
additional symmetries which are implied by the discrete symmetry in Table II.
In fact the terms in the superpotential
which are invariant under this discrete symmetry
are as well invariant under two global $U(1)$ symmetries.
The first one acts only on the color-triplet fields for which
the global $U_B(1)$ charges are respectively
$B(Q_i)=B(h_i)=+1/3$ and $B(u^c_i)=B(d^c_i)=B(h^c_i)=-1/3$.
$U_B(1)$ can be identified with Baryon number.
Under the second global $U_L(1)$ the color-singlet fields
transform with $L(L_\alpha)=L(H_3)=+1$,
$L(\nu^c_3)=L(S^c_3)=-1$ and  $L=0$ for the remaining
fields
$L_3$, $H_\alpha$, $H^c_i$, $\nu^c_\alpha$ and $S^c_\alpha$.
For the color-triplets the  L-charges are $L(h_i)=+1$,
$L(d^c_\alpha)=L(h^c_3)=-1$ and
$L(Q_i)=L(d^c_3)=L(h^c_\alpha)=0$.
$U_L(1)$ can be identified with Lepton number.
The $h_i$, $d^c_\alpha$ and $h^c_3$ heavy states which
carry both Baryon and Lepton numbers are leptoquarks.
B and L conservation in turn imply that R-parity is unbroken,
and then the model predicts a stable Lightest Supersymmetric Particle
(LPS).

{}From the assignments in Table II it is clear that most of the fields
acquiring a VEV transform non-trivially under the $Z_2\times Z_3$
symmetry. This is indeed unavoidable for any discrete
symmetry suitable for implementing the scheme which we have been discussing.
As a consequence, when the neutral components of the scalar fields
acquire a VEV, the discrete symmetry is spontaneously broken.
In the early universe, when a phase transition occurs during the
expansion, symmetry breaking takes place independently in different
causally disconnected regions that are filled with different
discrete phases, and separated from one another by domain walls.
In the standard hot universe theory, domain walls cause cosmological
problems since they would dominate the energy density
of the universe,  as well as astrophysical problems, since they
would lead to a considerable anisotropy in the
primordial background radiation [\cite{domain}].
Thus cosmological arguments would suggest that we have to renounce
the kind of models with spontaneously broken discrete symmetry
discussed here.
However, in an inflationary universe scenario it is possible to get rid
of this problem since inflation ensures that each region containing a
different phase becomes exponentially large, up to the point that
there would not be a single domain wall in the observable
part of the universe.
In order for this mechanism to be effective inflation has to go on
long enough after the phase transition, and the reheating temperature
of the universe after inflation should be low enough in order
to insure that the symmetry is not restored when the ordinary
adiabatic expansion of the standard cosmology begins.
Viable cosmological models in which
inflation takes place at the electroweak scale, and
that satisfy at these requirements exist, and
for example have been recently discussed in [\cite{ewinflation}].

\midinsert
{
\bs
$$
\vbox{\hsize= 13.3truecm
\baselineskip=12pt
\noindent TABLE II.
{\nine
Transformations of the fields in the three {\bf 27} of $\E$
under the discrete $Z_2\times Z_3$ symmetry. The index $i$ ranges from
1 to 3, while $\alpha=1,2$ refers to the first two generations.
}
\vskip -.8truecm}
$$
\def\om{\omit}
$$
\vbox{
\hsize= 13.35truecm
\offinterlineskip
\halign {
\vrule#&\strut\ $\hfil#\hfil$\ &\vrule#
      &&\strut\ $\hfil#\hfil$\ &\vrule#\cr
\noalign{\hrule}
height6pt &\om&\om&\om && \om&\om&\om&\om&\om & \cr
& \multispan3 \hfill $Z_2$ \hfill && \om&\om &Z_3& \om&\om & \cr
\noalign{\hrule}
height6pt & \om && \om && \om && \om && \om & \cr
& + && - && \ 1 \ && e^{-i{2\pi \o 3}} && e^{i{2\pi \o 3}} & \cr
\noalign{\hrule}
height6pt & \om && \om && \om && \om && \om & \cr
& [Q,u^c,H^c]_i && [e^c,h]_i && [u^c,h]_i && [d^c,h^c]_i && \om & \cr
height2pt & \om && \om && \om && \om && \om & \cr
& [H,h^c,\nu^c,S^c]_\alpha && [d^c,L]_\alpha && [\nu^c]_2 &&
[H^c]_\alpha \ [H^c,S^c]_2 && [H,\nu^c,S^c]_1 & \cr
height2pt & \om && \om && \om && \om && \om & \cr
& [d^c,L]_3 && [H,\nu^c,h^c,S^c]_3 && [Q,L,e^c,H^c]_3 && \om &&
[H,\nu^c,S^c]_3 & \cr
height4pt
& \om && \om && \om && \om && \om & \cr
\noalign{\hrule}}}
$$
\bs
}
\endinsert
\interlinea

\chap{phen} VI. Phenomenology.

We will now concentrate on the pattern of masses and mixings
allowed, in out scheme, for the light neutrinos.
We will first discuss the neutrino mass matrix in the one-loop
approximation, and then we will briefly describe the effects of the
additional contributions that arise from two-loop  diagrams.

\bigskip
\no{\subtitle One--loop neutrino masses. }
\medskip
\nobreak
In the previous section the general form
\req{2.9} of the one-loop neutrino mass matrix
was derived. The mass
terms for the light and hevy $Q=-1/3$ quarks given
in eqs. \req{2.11}-\req{2.14}
were also worked out, according to the choice of
the assignments in \req{2.10}.
Now from  Fig. 2,  we see that
in order to generate at one-loop the $\nu_\alpha$-$\nu^c_3$
mass term $[A_1B_2]$ in \req{2.9}, the two mass insertions
$[M_{\cal D}]_{i3}=\l4{\la\alpha\ra}i3\langle\tilde {S^c_\alpha}\rangle$
and $[m_{\cal D}]_{33}=\l2{\la1\ra}33\langle \tilde {N_1}\rangle$
are needed. This fixes $d_3$ and $h^c_3$ as the quarks that couple
to the external $\nu_\alpha$, implying
that the vertex $\l6\alpha33 \nu_\alpha Q_3 h^c_3$ must be
simultaneously non-vanishing.
On the other hand, in order to generate the  $\nu_\alpha$-$S^c_3$ mass
term $[A_1A_2]$, we need the mass insertion $[M_{\cal D}]_{i3}$
together with $[m_{\cal D}]_{\beta\gamma}=
\l6{\la3\ra}\beta\gamma\langle \tilde {\nu_3}\rangle$. This
in turn implies the non-vanishing of the
$\l6\alpha\beta3 \nu_\alpha Q_\beta h^c_3$ vertex.
However the two $\lw6$ vertices cannot be simultaneously
non-vanishing.
In fact this would require $Q_\beta$ to transforme like
$Q_3$ under the discrete symmetry, since they both couple to the same
bilinear $\nu_\alpha h^c_3$.
The result is that to this order only one of the two possible
radiative mass terms  for the $\nu_\alpha$
is allowed. A similar argument implies that also for the $N_3$
light state only one of the two $N_3$-$\nu^c_3$ and
$N_3$-$S^c_3$ one-loop mass terms is allowed, corresponding to
only one of the two $\l233\beta N_3 Q_3 d^c_\beta$ or
$\l23\alpha\beta N_3 Q_\alpha d^c_\beta$ being non-vanishing.

By requiring that our scheme should allow for non-zero
$\nu_\alpha$-$N_3$ mixings, we are left with the two choices {\it i)}
$\l4\alpha\beta3 \nu_\alpha Q_\beta h^c_3 = \l23\alpha\beta N_3
Q_\alpha d^c_\beta = 0$ or {\it ii)} $\l4\alpha33 \nu_\alpha Q_3
h^c_3 = \l233\beta N_3 Q_3 d^c_\beta = 0$.
The transformation properties listed in Table II correspond to the
first choice, and lead
to non-vanishing mass terms with $\nu^c_3$ while, at
this order, $S^c_3$ does not couple to any doublet neutrino and
remains massless.
As a result, at one loop the
Dirac mass matrix for the light neutrinos acquires the very simple form
$$
(\nu_1\>\nu_2\>N_3) \, \cdot {\cal M}_1 \cdot
\pmatrix{0 \cr S^c_3 \cr \nu^c_3 \cr},
\hskip 2.3 truecm
{\cal M}_1 =
\pmatrix{ 0&0&a_1 \cr 0&0&a_2 \cr 0&0&a_3 \cr}
\eq{3.1}
$$
where for convenience we have added one dummy entry
in the vector of the right handed neutrinos.
{}From \req{3.1} it is apparent that
two mass eigenstates $n_1$ and $n_2$
will be massless, while the third one
$n_3$ will acquire a Dirac mass $\mu_\nu=\sqrt{a_1^2+a_2^2+a_3^2}\,$.
Clearly here the $SU(2)$ singlet
$\nu^c_3$ plays the role of the right handed component of the massive
$SU(2)$ doublet neutrino.
As we will see in the following, the degeneracy between $n_1$ and
$n_2$ will be effectively removed due to two-loops corrections,
however for the moment we will discuss the results implied by
the mass matrix ${\cal M}_1$ in the one-loop approximation.

The unitarity transformation that relates the
flavor to the mass eigenstates is :
$$
\pmatrix{\nu_1 \cr \nu_2 \cr N_3 \cr} =
{\cal R} \pmatrix{n_1 \cr n_2 \cr n_3 \cr}
\qquad{\rm with}\qquad
{\cal R} =
\pmatrix{ c_\alpha          & 0       & s_\alpha \cr
          -s_\alpha s_\gamma & c_\gamma & c_\alpha s_\gamma \cr
          -s_\alpha c_\gamma & -s_\gamma & c_\alpha c_\gamma \cr }
\eq{3.2}
$$
where
$$
s_\alpha={a_1/\mu_\nu},              \qquad\qquad
c_\alpha s_\gamma={a_2/\mu_\nu},      \qquad\qquad
c_\alpha c_\gamma={a_3/\mu_\nu}.
$$
In \req{3.2} we have made use of the freedom in rotating the
degenerate massless states $n_1$ and $n_2$ in such a way that
$n_2$ does not couple to the electron
(we have implicitly assumed that the rotation of the
$e$ and  $\mu$ fields needed to diagonalize
their mass matrix has already been absorbed in the definition of
$\nu_1$ and $\nu_2$).

Following Ref.  [\cite{ellis-e6}],
in first approximation the mass entries in \req{3.1}
can be estimated to be
$$
\eqlabel{3.3}
\eqlabel{3.4}
\eqalignno{
&a_\alpha \sim
{1\o 32\pi^2} \l6\alpha33 \l73j3 m_b\, , \qquad\qquad\qquad \alpha=1,2
&\req{3.3}  \cr
&a_3 \sim
{1\o 32\pi^2} \l233\beta \l73j3 m_b.
&\req{3.4}  \cr
}
$$
\mn
Since all the three entries $a_1$, $a_2$ and $a_3$
in the mass matrix are proportional to the ``$b$"-quark mass
(which is the only light quark allowed to appear
inside the corresponding
loops) in principle there is no  reason to expect any
hierarchy among the $a_i$, and the mixing of the third generation
neutrino can then be large.
However, since $m_d$ and $m_s$  are both proportional to $\lw6 \la\tilde
{L_3}\ra$ and at the same time the ``$\tau$" mass is given by $\lw3
\la\tilde {L_3}\ra$, we would expect the $\lw6$ couplings, rather than
$\la\tilde {L_3}\ra$, to be small. On the other hand $m_b\sim \lw2
\la\tilde {H_1}\ra$ so that if we assume
$\la\tilde {L_3}\ra\sim \la\tilde {H_1}\ra$ and
that excessively large
hierarchies inside each set of $\lw m$ couplings are absent, it is
reasonable to expect that $\lw2 > \lw6$. In turn this implies $a_3 >
a_\alpha$ meaning that the light massive neutral state will be mainly the
third generation ``$\nu_\tau$" neutrino with $\mu_\nu \sim a_3$.


We can now question if $\mu_\nu$ can be small enough to
lie in a range of values
interesting for solving via matter enhanced
``$\nu_e$''--``$\nu_\tau$'' oscillations the solar neutrino problem.
As is well known a MSW solution would require
values of $\mu_\nu$ as small as $\lesssim 10^{-2}$ eV.
In order to estimate how small
$\mu_\nu$ can be for natural values of the
parameters,
we will assume the doublet VEVs to be $O(100)\,$GeV, so that in
\req{3.4} $\lw2 \sim m_d/100\,{\rm GeV} \simeq
10^{-4}$. Clearly we need as well a value for
$\l73j3$ since all the entries in ${\cal M}_1$ are proportional to this
coupling.
The set of $\lw7$ is responsible for generating
the masses for the new heavy states $h_j$-$d^c_\alpha$
through $\la\tilde {\nu^c_\alpha}\ra$, thus
their order of magnitude is in principle unknown.
However if we assume that the new
fermions and the new gauge bosons are as light as allowed by the present
phenomenological constraints (thus implying that they will be detected
with the next generation of colliders)
we can still work out an estimate for $\lw7$.
According to the present limits from direct searches at colliders,
the heavy fermions cannot be
much lighter than $\sim 100\,$GeV, so that
$\lw7\gtrsim 100{\rm GeV}/\la\tilde {\nu^c_\alpha}\ra$.
On the other hand we can argue that the lowest value
for this VEV is still expected to be $\sim O(10)\,$TeV.
By confronting the data on the light element abundances
with the standard nucleosynthesis calculations
a limit of 3.6 relativistic neutrinos in thermal
equilibrium at the time of nucleosynthesis can be derived
[\cite{nucleo-nu}].
This implies that not even one additional neutrino
can remain in equilibrium in addition to the three known light states.
Though these are singlets under $SU(2)\times U(1)$, the two additional light
neutrinos present in our scheme do have
$U(1)^\pr$ interactions. Therefore, we have to require this interaction
to be weak enough to allow for the decoupling of both the $SU(2)$ singlets
$\nu_3^c$ and $S^c_3$
at a sufficiently early time (say, before the QCD phase transition)
so that their number density
can be safely diluted.
This argument implies that
the mass of the lightest additional gauge boson $M_{\beta}$ should be
at least of the order of $\sim 1$-$2\,$TeV [\cite{nucleo-e6}].
Such a large mass will be mainly generated by the singlet VEVs
giving $M_\beta \sim g_\beta [(Q_\beta^{\nu^c}\la\tilde
{\nu^c_\alpha}\ra)^2
+(Q_\beta^{S^c}\la \tilde {S^c_\alpha}\ra)^2]^{1/2}$.
Since the coupling constant
$g_\beta\sim g_1 \sim 0.16$ and the $Q_\beta$ charges $\lesssim1$, we see
that indeed VEVs $\sim 10$ TeV are required, implying as a result
$\lw7 \sim 10^{-2}$.
We also note that since the only non-vanishing Yukawa coupling
for $\nu^c_3$ is precisely $\l73j3$ we could have also attempted to
estimate it directly by requiring that the
exchange of scalar quarks
of a typical SUSY mass $\sim$ few 100 GeV should not be able to
keep this particular species in thermal equilibrium.
As a result of a  rough computation we have found that
scalar quark masses below 1 TeV are indeed
consistent with $\l73j3 \sim 10^{-2}$.

Now, according to \req{3.3}
the order of magnitude of the ``$\nu_\tau$" mass is $\mu_\nu \sim 10\,$eV.
This value is indeed too large to play any role
in the solar (or atmospheric) neutrino problem, however
it is not in conflict with the cosmological limit
$\mu_\nu \lesssim 92\,\Omega\, h^2\,$eV implied by the requirement
not to over-close the universe.
Since this neutrino is effectively stable
it could be a natural candidate for the hot component
of the dark matter. We note as well that
our scheme is also consistent with a certain amount of cold
dark matter, since
$R$-parity is unbroken and the LPS is stable.

On the other hand if  $\l73j3$ were about two order of magnitude
smaller, then the $n_3$ mass would fall in the right range of values
for a possible explanation of the atmospheric neutrino
deficit via ``$\nu_\mu$''--``$\nu_\tau$'' oscillations. We will
briefly discuss in the following a possible scheme in which
the fact that the two flavor states do have
different NC interactions could play an interesting role
for these oscillations.

\bigskip
\no{\subtitle Two--loop neutrino masses. }
\medskip
\nobreak
The previous discussion does not imply that the scenario that we are
analyzing cannot offer a solution to the solar neutrino problem.
In fact even if at the one loop level
both ``$\nu_e$" and ``$\nu_\mu$'' are massless, non zero
$\nu_\alpha$-$S^c_3$ entries can be generated at the two loop
level due to the presence of the $\lw5\lc$ couplings in $W_2$
\req{2.4}.
A typical two-loop diagram is depicted in Fig. 3.
We note that since the set of couplings needed to generate this
diagram is indeed allowed by the assignments in Table II,
the generation of two-loops mass entries is not in
conflict with the other constraints on the superpotential
discussed in the previous section.
\midinsert
{
\bs
\epsfxsize=13.3truecm 
\epsffile{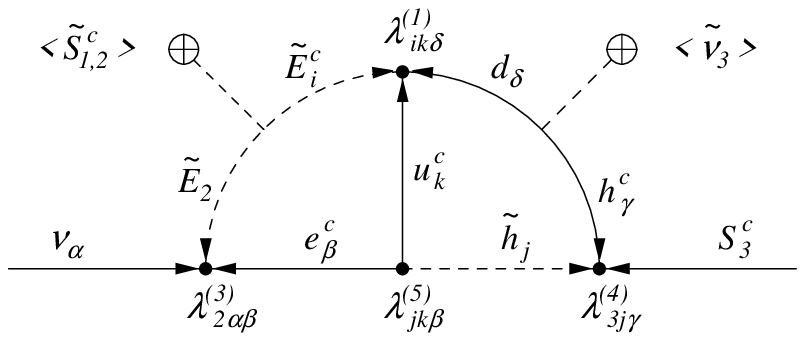}
$$
\vbox{\hsize= 13.3truecm
\baselineskip=14pt
\noindent Fig. 3:
{\nine
A two-loop diagram giving rise to $\nu_\alpha$-$S^c_3$ entries
in the neutrino Dirac mass matrix \req{3.2}.}
}
$$
}
\endinsert
\interlinea
\noindent
At the next order two additional entries
are generated in
the neutrino Dirac mass matrix ${\cal M}_2$, namely
$[{\cal M}_2]_{12}\equiv b_1$ and
$[{\cal M}_2]_{22}\equiv b_2$.
They can be
roughly estimated to be [\cite{ellis-e6}].
$$
b_\alpha \sim
\l1i k\delta \l32\alpha\beta\l43j\gamma \l5j k\beta
{m_s\o (16 \pi^2)^2}.
\eq{3.5}
$$
We note that in
the present case the ``$s$"-quark $(d_2 h^c_2)$
is the heaviest one allowed to run
inside the loop.
Now if we take $\l321\beta \sim {m_{e} /100\, {\rm GeV}}$,
$\l322\beta \sim {m_{\mu} /100\, {\rm GeV}}$ and
the representative value
$\lw1 \sim m_c/ 100\, {\rm GeV}$,
we obtain $b_2 \sim
10^{-1} \lw4\lw5 $eV $\gg b_1$ showing that values of the
``$\nu_\mu$" mass interesting for a MSW [\cite{MSW}]
solution of the solar neutrino
problem can indeed be accommodated for natural values of the remaining
Yukawas.

To summarize our results, in this model we have a massive
$n_3$ neutrino, mainly ``$\nu_\tau$", with a mass
that can easily fall in the range
$\sim 0.1-10\,$eV interesting
for providing a hot DM candidate or
for the oscillations of the ``\num'' atmospheric neutrinos.
A second neutrino $n_2$, mainly ``$\nu_\mu$",
acquires a much smaller mass at the two
loop level and can be relevant for matter enhanced ``\nue''-``\num''
oscillations in the sun and finally,
due to the absence in our minimal scheme of
a helicity partner, $n_1$ remains massless.

\bigskip
\no{\subtitle A scheme with a light Z$_{\bf \beta}$ boson. }
\medskip
\nobreak
Before concluding this section we want to illustrate
a different scheme in which an attempt
for an unconventional solution to the solar neutrino
problem can be formulated, and in which unusual effects
for the ``$\nu_\mu$''--``$\nu_\tau$'' oscillations
of the atmospheric neutrinos could arise as well.
The following analysis
is mainly intended as an example
of the unusual phenomenology that can be implied by
UA models.

{}From the previous discussion it should be clear that if
we insist on trying to
achieve a very small one-loop $\mu_\nu$ mass,
we must require $\la\tilde {\nu^c_\alpha}\ra$
to be very large in order to allow for
$\lw7 \ll 10^{-2}$ while still keeping the vectorlike quarks
as heavy as $\gtrsim 100\,$GeV.
Then let us assume that
$\la\tilde {\nu^c_\alpha}\ra$
is much larger than all the other VEVs
including
$\la\tilde {S^c_\alpha}\ra$.
With this assumption the
$Z_\psi$--$Z_\chi$ mixing angle is

$$
\tan 2\beta = {2 \sum_j Q^j_\psi Q^j_\chi
\la\phi_j\ra^2 \o
\sum_j (Q^j_\psi)^2 \la \phi_j \ra^2 -
\sum_j (Q^j_\chi)^2 \la \phi_j \ra^2 }
\simeq
-{\sqrt{15}\o 7}
\eq{3.6}
$$
\mn
where the $\la \phi_j \ra$'s represent the various singlet and doublet VEVs
occurring in the model.
Then the first gauge boson in \req{2.3}
$Z_\beta^\pr=(-1/4)Z_\psi + (\sqrt{15}/4)Z_\chi$
is very heavy.
The second one $Z_\beta^{\pr\pr}$,
which corresponds to the orthogonal combination of generators,
does not couple to the $\nu^c$ states.
A major consequence of this fact is that, as
far as the remaining VEVs which contribute to its mass are not
very large, the $Z_\beta^{\pr\pr}$ boson will be light.
A second consequence is that the gauge interactions
cannot keep the the right handed light $\nu^c_3$ neutrino
in thermal equilibrium in the early universe,
since the $\nu^c$'s are effectively singlets with respect to all the
`light' gauge bosons.
Furthermore since now $\l73j3 \ll 10^{-2}$, even the exchange of
scalar quarks as light as $\sim 100\,$GeV cannot help in thermalizing
these light states, so that the
$\nu^c_3$ degree of freedom will not be populated at the time of
nucleosynthesis.
Clearly the other singlet $S^c_3$
will still be coupled to the light $Z^{\pr\pr}_\beta$
since there is no possible choice for the angle $\beta$ for which
both the states are decoupled.
However, in contrast to the previous case in which the presence of a light
$S^c_3$ was needed in order to generate two loop masses, we will now
assume that $S^c_3$ transforms under the discrete symmetry like
one of the $S^c_\alpha$ so that it will acquire a large mass.
We obtain here a truly `minimal' scheme with only one
light singlet neutrino. However, in contrast to a similar scheme
first proposed in [\cite{cecilia}] there are no Majorana entries in the mass
matrix. We stress that in this second scheme
the requirement of allowing for a very small ``$\nu_\tau$''  mass $\mu_\nu
\ll 1\,$ eV automatically allows for a new `light' neutral gauge
boson as well.

Since the nucleosynthesis constraints on the mass of the lightest
additional gauge boson $Z_\beta^{\pr\pr}$ are evaded, this boson
could be as light as allowed by the present limits from direct
searches at colliders [\cite{zp-direct}] and from the analysis of
$Z^\pr$ indirect effects [\cite{zp-new}], resulting in both cases in
$M_\beta \gtrsim 200\,$GeV.
One might object that these limits cannot be straightforwardly
applied to the present situation, since they are derived from
analyses based on the conventional scheme, while in the
present case a large number of fermion couplings are clearly
different. However, since we have no reason to expect that in the UA schemes
the bounds could be greatly strengthened or relaxed, we will assume that the
quoted limit still holds also here.

The presence of a light $Z_\beta$ is crucial for the following
discussion. In fact, as
we have already stressed, the UA scheme implies
that the ``$\tau$'' neutrino does not have the same
$U(1)^\pr$ interactions with respect to ``\nue'' and ``\num''.
It is then interesting to study the implications of having
the different neutrino flavors interacting differently with
matter through NC.
For the present discussion we will restrict ourself to the
two flavor case $\nu_\alpha$ ($\alpha=1$ or $2$) and $N_3$.

The propagation of the two neutrino flavor
eigenstates through matter is governed by the
Schr\"odinger-like
time evolution equation  [\cite{MSW}]
$$
i{d\o dx}{\nu_\alpha\choose N_3} = {1\o 2E} \,
{\cal H} \, {\nu_\alpha\choose N_3}
\eq{3.7}
$$
The effective Hamiltonian in \req{3.7} relevant to the present case
is
$$
{\cal H} ={1\o 2}{\cal R}_{\alpha}
\pmatrix{-\mu_\nu^2 & 0 \cr  0  & \mu_\nu^2}
     {\cal R}_{\alpha}^\dagger +
2E  \pmatrix{{\cal A}_{CC} - \Delta_{NC} & 0 \cr
    0 &  - {\cal A}_{CC} + \Delta_{NC} \cr}.
\eq{3.8}
$$
In \req{3.8} $\mu_\nu$ is the $n_3$ mass while
${\cal R}_{\alpha}$ is the relevant 2$\times$2 \ $\nu_\alpha$--$N_3$
vacuum mixing matrix.
${\cal A}_{CC}$ represents the
coherent neutrino forward scattering due to the Charged Current (CC)
interaction. Then in
the case of the solar electron-neutrinos ($\alpha=1$)
${\cal A}_{CC}=\sqrt{2}G_F {\cal N}_e$
with ${\cal N}_e$ the electron density in the sun, while
for the upward-going
atmospheric ``\num'' neutrinos propagating through the earth
($\alpha=2$) ${\cal A}_{CC}= 0$.
Finally
$\Delta_{NC}$ represents the difference in the
forward scattering between $\nu_\alpha$ and $N_3$
that is due to the difference in
their NC interactions. Clearly, due to universality,
 this term vanishes exactly in
the SM. However in the present case it is non-zero
due to the additional contribution from $Z^{\pr\pr}_\beta$ exchange.
The additional term can be written as
$$
\eqalign{
\Delta_{NC} &= 2 \sqrt{2} G_F {M^2_{Z_0}\o M^2_\beta}
{g^2_\beta\o g^2_0}\> {\cal F}(Q_\beta) \>             \cr
{\cal F}(Q_\beta) &=
(Q_\beta^{\nu_\alpha} - Q_\beta^{N_3})
(Q_\beta^e + Q_\beta^p + Y_n Q_\beta^n) {\cal N}_e \cr }
\eq{3.9}
$$
where ${g^2_\beta/g^2_0}$ in the first equation
is the ratio of the squared $U_\beta(1)$
and $SU(2)$ gauge coupling constants.
In the expression for ${\cal F}(Q_\beta)$,
$Q_\beta^f\equiv Q_\beta(f) - Q_\beta(f^c)\>$
is the vector coupling of the
$f=e,p,n$ fermion to the $Z^{\pr\pr}_\beta$ boson.
We have taken for the proton density
${\cal N}_p={\cal N}_e$ corresponding to
electrically  neutral matter, and finally
$Y_n$ is the ratio of the neutron to electron
density.
We note that in contrast to the SM, here the NC forward scattering off
electrons and off protons do not cancel.
Since both the $u$ and $u^c$ quarks belong to the {\bf 10}
of $SU(5)$ (see Table I) the $Q_\beta^u$ vector charge vanishes, and the
contribution of the
scattering off nucleons is determined only by the ``$d$'' quark density.
We also note that for the present case, corresponding to
$\sin\beta =-1/4$,
the light $Z^{\pr\pr}_\beta$ is mainly a $Z_\psi$ boson.
Had we chosen for the
``$d$"-quarks instead of the assignments \req{2.10}
the alternative assignments
$``d^c_\alpha"\equiv d^c_\alpha \in {\bf 16}$ and
$``d^c_3" \equiv h^c_3 \in {\bf 10}$,
the $Q_\psi^d$ vector charges of the $d$ quark would have been
zero, as is the case for $Q_\psi^e$,  and the
$\Delta_{NC}$ term would have been accordingly suppressed.

The presence of interactions implies that the
``in matter'' mass eigenstates are different from the vacuum
mass eigenstates [\cite{MSW}]. They can be obtained
by diagonalizing ${\cal H}$ in \req{3.8}.
The two eigenvalues $\pm\delta$
and the matter mixing angle $\alpha_m$ are
given respectively by
$$
\eqalign{
&\delta^2=\big[{4E\o \mu_\nu^2}({\cal A}_{CC} - \Delta_{NC}) -
\cos2\alpha\big]^2 + \sin^2 2\alpha   \cr
&\sin^22\alpha_m=\sin^22\alpha/ \delta^2.
}\eq{3.10}
$$
The second equation in \req{3.10} shows
that if the vacuum mixing angle is close to maximal
($\sin2\alpha\simeq1$) the effect of the additional interactions
would be that of {\it reducing} the mixing in matter
by the factor $4E({\cal A}_{CC} - \Delta_{NC})/\mu_\nu^2\,$,
thus suppressing the oscillation of the high energy neutrinos
with respect to the low energy ones.
If in contrast $\sin2\alpha$ is small, the mixing in matter
will be maximal
in the resonance region defined by
$$
{\mu_\nu^2 \o 4E}\cos2\alpha = {\cal A}_{CC} - \Delta_{NC}.
\eq{3.11}
$$
We see that as far as the second term in the l.h.s. is not
completely negligible with respect to the first one,
the allowed regions in the $(\Delta m^2, \sin^2 2\theta)$
plane would be different than in the standard case.

Now by confronting \req{3.9} with the electron-neutrino  CC
forward scattering  we obtain that the l.h.s. in \req{3.11}
would vanish for
$$
{\Delta_{NC} \o {\cal A}_{CC} } =
2{\cal F}(Q_\beta) {M^2_{Z_0}\o M^2_\beta} {g^2_\beta\o g^2_0}=1,
\eq{3.12}
$$
\ie when the difference between the $\nu_1$ and $N_3$ NC interactions
compensates in full the $\nu_1$ CC interaction.
Clearly in this case we would have found the possibility of a
$\nu_1$--$N_3$ resonant conversion
even in the case of practically massless neutrinos
($\mu_\nu \ll 10^{-3}$).
In the present case, in which
the requirement
of a large $\la\nu_\alpha^c\ra$ and
a $\nu^c_3$ decoupled from the light $Z_\beta^{\pr\pr}$
selects $\sin\beta=-1/4$ and by taking
$Y_n=0.5$ as the maximum value of
the neutron density at the center of the sun,
the term ${\cal F}(Q_\beta)$ in \req{3.12}
gives an enhancement factor $\sim 2$.
However,
with the normalizations given in Table I,
the ratio of the squared $U_\beta(1)$
and $SU(2)$ coupling constants
is of the order of the electroweak mixing angle
$\sin^2\theta_w \sim 1/4$ and then we see that
the massless neutrino case is indeed ruled out,
since it would require $M_\beta \sim M_{Z_0}$.

In the case of ``\num'' propagation trough matter,
the same mechanism could affect the rate of
$\nu_2$-$N_3$
conversion for the upward-going
atmospheric muon-neutrinos.
The Kamiokande II [\cite{KAM92a}] and IMB [\cite{IMB92a}]
collaborations have observed an anomaly in the ratio of muons to
electrons events induced by atmospheric neutrinos of energies of a few
hundreds MeV, and a possible explanation of the effects has been given
in terms of $\nu_\mu \rightarrow \nu_x$ oscillations where
$\nu_x=\nu_e$, $\nu_\tau$ or a sterile neutrino ($\nu_s$).
For explaining the data the
$\nu_\mu$--$\nu_x$ mixing angle is required to be close to
maximal ($\sin^2 2\theta > 0.5$, see \eg [\cite{KAM92c}]).
However the IMB [\cite{IMB92b}] and Baksan [\cite{Baksan91}] experiments have
observed no reduction for the $\nu_\mu$ flux of upward-going neutrinos
with $E\gtrsim 1$-$2\,$GeV,
and these data have been used to set stringent limits on
the allowed region in the $(\Delta m^2, \sin^2 2\theta)$ plane
[\cite{zurab}].

As we have already said, for large $\sin 2\theta$
the effect of possible additional interactions with matter would in general be
that of shifting the matter mixing angle away from maximal, thus {\it
suppressing} the rate of conversion.
The equations describing this case would still be \req{3.10} and
\req{3.11}
with the correct values of the electron, proton and neutron densities
in the earth
and with ${\cal A}_{CC}=0$ for $\nu_x\neq \nu_e$.
While negligible at low energy,
the effect of the interaction with matter could become particularly
relevant for high energy neutrinos, thus helping to explain the
data.

Such a mechanism has been investigated in Ref. [\cite{atmnu-matter}]
for the case of $\nu_\mu$ oscillating into electron or sterile
neutrinos. In particular for $\nu_\mu$--$\nu_s$ oscillations the
difference in the interaction strength of the two neutrino species is
about 1/2 that of the standard $\nu_e$ CC interaction, and the
analysis in [\cite{atmnu-matter}] shows that indeed in this case matter
effects are important.
In our case,  due to the difference in the $U_\beta(1)$ charges,
similar effects could arise for ``\num''--``\nut'' oscillations as
well.
For the propagation through the earth $Y_n\simeq 1$
gives ${\cal F}(Q_\beta)\sim 2.7$, and we can assume,
consistently with the present
direct [\cite{zp-direct}] and indirect [\cite{zp-new}]
limits, $M_\beta\sim 200\,$GeV.
Then from the l.h.s. of \req{3.12} we see that
the effective strength of the new interaction relative to the standard CC
interaction is $\sim 0.27$, showing that
in this case sizeable effects could be present as well.

\chap{Conc} V. Conclusions.

In conclusion we have examined the possibility of constructing
consistent models in which the known fermions of the three different
generations do not have the same gauge interactions under possible
additional $U(1)^\pr$ factors. We have carried out our analysis in the
frame of the superstring--inspired $\E$ models, taking as a guideline
for constructing our scheme, the requirement of having interesting
neutrino phenomenology with naturally small radiatively generated
Dirac masses. We have shown that models based on this scheme are
indeed viable. They can be realised by imposing a family-non-blind
discrete symmetry on the superpotential.
We have discussed in some detail a minimal model, in which only two
additional light $SU(2)$ singlet neutrinos are present thus leaving
one doublet neutrino massless.
Clearly other models based on the same scenario but
with a more rich structure in the neutrino sector can
also be constructed.

We have shown that, in our model, values of
the neutrino masses in interesting ranges for explaining
the solar and atmospheric neutrino anomalies,
or possibly for providing a hot component of the DM,
can be obtained with a natural choice of the parameters.
In addition, since baryon and lepton numbers
are both conserved,  the proton is effectively stable.
Due to the presence of  FCNC in the up-quark sector,
a rate for $D^0$--$\bar {D^0}$ oscillations larger than in the SM,
but still consistent with the present limits, is predicted.
However there are no other dangerous sources of FCNC in the model.

In order to illustrate some unusual consequences of our scheme,
we have also investigated a different scenario in which only one
$SU(2)$ singlet neutrino is light, and an additional neutral gauge boson
is allowed at energies as low as $\sim 200\,$ GeV.
Since the additional neutrino naturally decouples
from the new light gauge boson,
there is no conflict with the nucleosynthesis constraints
on the number of neutrino species.
We have shown that the generation-dependent NC interaction mediated by
this gauge boson, though probably not relevant in the case of
the propagation of the
solar electron-neutrinos through matter, could however be of
some importance in the case
of $\nu_\mu$--$\nu_\tau$ oscillations for the upward-going
atmospheric neutrinos.

\chap{Ackn} Acknowledgements.

We wish to acknowledge
several discussions with R. Akhoury, A. Dolgov, G. Kane and I.
Rothstein, which have been of primary importance for the realization
of this work, and we thank F. del Aguila for a critical reading
of the manuscript.
We also thank Z. Berezhiani for pointing out the implications of
having different $\nu_\mu$ and $\nu_\tau$ NC interactions for the
upward-going atmospheric neutrinos, and A. Kassa for a valuable help
with the graphics.

\vfil\eject

\vskip4truecm

\null
\baselineskip 8pt
\centerline{\title References}
\vskip .8truecm

\biblitem{see-saw}
M. Gell-Mann, P. Ramond, and R. Slansky, in {\it
Supergravity}, F. van Nieuwenhuizen and D. Freedman eds., (North
Holland, Amsterdam, 1979) p.~315; \hbup
T. Yanagida, {\it Proc. of the
Workshop on Unified Theory and Baryon Number of the Universe}, KEK,
Japan, 1979.  \par

\biblitem{rizzo-e6}
For a review see J.L. Hewett and T.G. Rizzo, \prep{183} 195 (1989).
\par

\biblitem{ellis-e6}
B.A. Campbell \ea, \ijmpa{2} 831 (1987). \par

\biblitem{MNS}
A. Masiero, D.V. Nanopoulos and A.I. Sanda,
\prl{57} 663 (1986).  \par

\biblitem{branco}
G.C. Branco and C.Q. Geng, \prl{58} 969 (1986).\par

\biblitem{witten-yuk}
E. Witten, \npb{258} 75 (1985). \par

\biblitem{MSW}
L.~Wolfenstein, \prd{17} 2369 {1978}; {\bf D20}, 2634 (1979); \hbup
S.~P.~Mikheyev and A.~Yu.~Smirnov,
{\it Yad.  Fiz.} {\bf 42}, 1441 (1985); \ncim{{9C}} 17 (1986).\par

\biblitem{e6-super}
D. Gross, J. Harvey, E. Martinec and R. Rhom, \prl{54} 502 (1985);
\npb{256} 253 (1985); \ib {\bf B 267} (1986) 75; \hbup
P. Candelas, G. Horowitz, A. Strominger and E. Witten, \npb{258}
46 (1985); E. Witten, \plb{149} 351 (1984). \par

\biblitem{weinberg-fc}
S. Glashow and S. Weinberg, \prd{15} 1958 (1977).\par

\biblitem{ll1}
P. Langacker and D. London, \prd{38} 886 (1988). \par

\biblitem{lfc}
E. Nardi, \prd{48}, 1240 (1993). \par

\biblitem{domain}
Ya. B. Zeldovich, I. YU. Kobzarev and L. B. Okun, \plb{50} 340
(1974).\par

\biblitem{ewinflation}
L. Knox and M. S. Turner, \prl{70} 371 (1992). \par

\biblitem{nucleo-nu}
T. Walker, in  Texas/PASCOS 92:
{\it Relativistic Astrophysics and Particle Cosmology, Proceedings of
the 16th Texas Symposium on Relativistic Astrophysics and 3rd
Particles, Strings, and Cosmology Symposium}, Berkeley, CA, 1992,
edited by C.W. Akerlof and M.A. Srednicki (The New York Academy of
Sciences, vol. 688, New York, 1993), p. 745.  \par

\biblitem{nucleo-e6}
M.C. Gonzales-Garcia and J.W. Valle, \plb{240} 163 (1990); \hbup
J. L Lopez and D.V. Nanopoulos, \plb{241} 392 (1990); \hbup
A.E. Faraggi and D.V. Nanopoulos, Mod. Phys. Lett. {\bf A6} 61 (1991). \par

\biblitem{cecilia}
C. Jarlskog, Nucl. Phys. {\bf A518}, 129 (1990). \par

\biblitem{zp-direct}
CDF Collaboration, F. Abe \ea, \prl{67} 2609 (1991); \ib {\bf 68},
1463 (1992). \par

\biblitem{zp-new}
E. Nardi, E. Roulet and D. Tommasini, \prd{46} 3040 (1992); \hbup
P. Langacker and M. Luo, \prd{45} 278 (1992); \hbup
F. del Aguila, W. Hollik, J.M. Moreno and M. Quir\'os, \npb{372}
 3 (1992); \hbup
J. Layssac, F.M. Renard and C. Verzegnassi, \zpc{53} 97 (1992); \hbup
M.C. Gonzalez Garc\'\i a and J.W.F. Valle; \plb{259} 365 (1991); \hbup
G. Altarelli et al., \plb{263} 459 (1991). \par

\biblitem{atmnu-matter}
E. Akhmedov, P. Lipari and M. Lusignoli; University ``La Sapienza" -
Rome Report No. 912 (November 1992). \par

\biblitem{KAM92a}
K.S. Hirata  {\it et al.} (Kam-II Collaboration),
\plb{280} 146 (1992). \par

\biblitem{IMB92a}
R. Becker-Szendy {\it et al.} (IMB Collaboration), \prd{46} 3720
(1992); see also D. Casper {\it et al.}, \prl{66} 2561 (1991). \par

\biblitem{KAM92c} E.W. Beier  \ea \ \plb{283} 446 (1992). \par

\biblitem{IMB92b} R. Becker-Szendy {\it et al.} (IMB Collaboration),
\prl{69} 1010 (1992). \par

\biblitem{Baksan91}
Baksan Collaboration,
M.M. Boliev {\it et al.}
in {\it Proceedings of the 3rd International Workshop on
Neutrino Telescopes,} Venice, Italy, 1991, edited by Milla Baldo Ceolin
(Istituto Nazionale di Fisica Nucleare, Padova, 1991), p. 235 \par

\biblitem{zurab}
Z. Berezhiani, in
{\it Proceedings of the International Workshop on the $\nu_e/\nu_\mu$
problem in Atmospheric Neutrinos}, Gran Sasso, Italy, 1993,
edited by V. Berezinsky
(Istituto Nazionale di Fisica Nucleare, Gran Sasso, 1993). \par

\interlinea

\insertbibliografia

\vfil \eject \bye